\documentstyle[12pt]{article}
\newcommand\mysection{\setcounter{equation}{0}\section}

\newskip\humongous \humongous=0pt plus 1000pt minus 
1000pt

\newif\ifdtup

\def\figcap{\section*{Figure Captions\markboth
        {FIGURECAPTIONS}{FIGURECAPTIONS}}\list
        {Figure 
\arabic{enumi}:\hfill}{\settowidth\labelwidth{Figure
999:}
        \leftmargin\labelwidth
        \advance\leftmargin\labelsep\usecounter{enumi}}}
 \relax
\def\MC{Monte Carlo}
\def\s#1{{\small #1}}
\def\DL{\s{DL}}
\def\LO{\s{LO}}
\def\NLO{\s{NLO}}
\def\EV{\s{EVENT}}
\def\bom#1{{\mbox{\boldmath $#1$}}}
\def\beq#1{\begin{equation}\label{#1}}
\def\beeq#1{\begin{eqnarray}\label{#1}}
\def\const{\mbox{const.}}
\def\eeq{\end{equation}}
\def\eeeq{\end{eqnarray}}
\def\as{\alpha_S}
\def\asb{\bar\alpha_S}

\def\ee{e^+e^-}
\def\Clim{\raisebox{-1ex}{\rlap{\tiny $\;\;C\to 0$}} 
\raisebox{0ex}{$\;\;\;\,\sim\;\;\;\,$}}
\def\Climeq{\raisebox{-1.2ex}{\rlap{\tiny $\;\;C \sim 3/4$}} 
\raisebox{0ex}{$\;\;\;\,\simeq\;\;\;\,$}}
\def\Cpar{$C$-parameter}
\def\frac#1#2{ {{#1} \over {#2} }}
\def\rat#1#2{\mbox{\small $\frac{#1}{#2}$}}
\def\half{\rat 1 2}
\def\thrd{\rat 1 3}
\def\thrq{\rat 3 4}
\def\twth{\rat 2 3}

\def\kper{k_{\perp}}
\def\smin{s_{\mbox{\scriptsize min}}}

\def\cav#1{Cambridge preprint Cavendish--HEP--#1}

\def\np#1#2#3{Nucl.\ Phys.\ B#1 (19#3) #2}
\def\pl#1#2#3{Phys.\ Lett.\ #1B (19#3) #2}
\def\pr#1#2#3{Phys.\ Rev.\ D #1 (19#3) #2}

\def\prep#1#2#3{Phys.\ Rep.\ #1 (19#3) #2}
\def\prl#1#2#3{Phys.\ Rev.\ Lett.\ #1 (19#3) #2}

\begin{document}
\begin{titlepage}
\renewcommand{\thefootnote}{\fnsymbol{footnote}}
\begin{flushright}
     Cavendish--HEP--97/10 \\
     LPTHE--ORSAY 97/39 \\
     September 1997 \\
     hep-ph/9710333
\end{flushright}
\vspace*{\fill}
\begin{center}
{\Large \bf Infrared Safe but Infinite: \\[1ex]
Soft-Gluon Divergences Inside the Physical
Region\footnote{Research supported in part by
the U.K. Science and Engineering Research Council
and by the EC Programme ``Human Capital and Mobility", Network 
``Physics at
High Energy Colliders", contract CHRX-CT93-0357 (DG 12 
COMA).}}
\end{center}
\par \vskip 2mm
\begin{center}
        {\bf S.\ Catani} \\
        INFN and Dipartimento di Fisica, Universit\`a di Firenze,\\
        Largo E. Fermi 2, I-50125 Florence, Italy\\
        LPTHE, Universit\'{e} Paris-Sud, B\^{a}timent 211,
        F-91405 Orsay, France
        \par \noindent
        and
        \par \noindent
        {\bf B.R.\ Webber} \\
        Cavendish Laboratory, University of Cambridge,\\
        Madingley Road, Cambridge CB3 0HE, U.K.
\end{center}
\par \vskip 2mm
\begin{center} {\large \bf Abstract} \end{center}
\begin{quote}
We show that QCD observables defined as infrared- and collinear-safe,
according to the usual Sterman-Weinberg criteria, can nevertheless
be infinite at accessible points inside phase space, to
any finite order of perturbation theory.  Infrared finiteness is
restored after resummation of divergent terms to all orders.  The
resulting characteristic structure, which we call a {\em Sudakov
shoulder}, represents an interesting new class of QCD predictions. 
\end{quote}
\vspace*{\fill}
\end{titlepage}
\pagestyle{plain}
\renewcommand{\thefootnote}{\fnsymbol{footnote}}
\mysection{Introduction}
\label{intro}

The perturbative QCD approach [\ref{book}] to the calculation of hadronic 
cross-sections at large momentum transfer $Q$ is based on the 
Sterman-Weinberg criteria [\ref{SW}] of infrared and collinear {\em safety},
combined with the factorization theorem for mass singularities [\ref{CSS}].
A hadronic observable is said to be infrared and collinear safe 
if it is insensitive to the emission of soft momenta and to the
splitting of a final-state momentum into collinear momenta.
The properties of 
infrared and collinear safety guarantee that the long-distance
component of the scattering process, which is controlled by non-perturbative
phenomena, is suppressed by some inverse power of $Q$ as the hard scale $Q$ 
increases. Thus, the observable is dominated by short distances
and the short-distance component, which depends only logarithmically on
$Q$, is computable as a power-series expansion in the strong
coupling $\as(Q)$.

It is commonly assumed that the consequences of the Sterman-Weinberg
criteria are even wider:
the power-series in $\as$ is assumed to have coefficients
that are finite order by order in perturbation theory.
Interpreted in this way, the criteria become a statement on the
infrared and collinear {\em finiteness} of
the fixed-order perturbative expansion.

In this paper we would like to point out that, in general, this strong
interpretation of the criteria is not valid. In fact, the perturbative
expansions of infrared- and collinear-safe quantities can be divergent
{\em order by order} in perturbation theory and these divergences are
still produced by the radiation of soft and/or collinear partons. 

The presence of logarithmic singularities 
that spoil the convergence of the perturbative expansion near the exclusive
boundaries of the phase space of infrared- and collinear-safe observables
is well known [\ref{CTWT}-\ref{CMNT}]. These singularities reflect the 
difficulty in extending the Sterman-Weinberg criteria towards extreme 
kinematical regions and, in many cases, they can be handled by all-order 
resummation methods.

The main issue to be discussed in the present paper concerns a
new general class of singularities that appear {\em inside the physical
region} of the phase space. They arise whenever the observable in question
has a non-smooth behaviour in some order of perturbation theory at an
accessible point, which we shall call a {\em critical point},
inside phase space. This can happen if the phase-space
boundary for a certain number of partons lies inside that for a larger
number, or if the observable itself is defined in a non-smooth way. In
either case, if the distribution of the observable is actually
discontinuous at that point in some order, it will become infinite
there to all (finite) higher orders.  If it is continuous but
not smooth, i.e.\ if some derivatives are discontinuous, then
the higher-order predictions will become unstable, with infinite
derivatives, at that point. 

There are several possible reactions to this state of affairs. One can try
to avoid non-smooth observables altogether. As we shall see, this actually
rules out some commonly-used quantities. One can use such observables
but avoid the phase-space regions near critical points. In this
case, one still needs some understanding of the phenomenon in
order to assess the extent of the dangerous regions. Alternatively,
one can identify the problematic terms in each order of perturbation
theory and resum them to all orders.  As mentioned above,
this approach has been used successfully for the treatment of
singularities near the exclusive phase-space boundary in many
observables. We advocate a similar approach to the new class
of singularities discussed here, although much work remains
to be done in order to reach a comparable level of understanding.
As a first step in this direction, we carry out the resummation
of leading double-logarithmic terms, and show that it leads to
smooth behaviour at the critical point.

In the following Section we give a general discussion of the effects of
soft and/or collinear parton emission in fixed-order QCD perturbation
theory. We first review the cancellation mechanism which normally ensures
the regularity of infrared- and collinear-safe observables, and the way
in which this breaks down at the exclusive boundary of phase space.
We then show how a similar breakdown will occur inside phase space
when the observable is defined in a non-smooth way, and describe
the general form of the resulting singularities.

In Sect.~\ref{exam} we discuss in some detail two specific cases in which
such singularities actually appear. The first is the distribution
of the \Cpar\ [\ref{Cpar},\ref{CWlett}], 
a well-known event shape variable for
$\ee$ annihilation final states. The physical region is $0<C<1$,
but at the point $C=3/4$ there is a divergence
in the perturbative prediction of the $C$-distribution at second
and higher orders.  The second example is taken from jet physics
in hadron-hadron collisions, and concerns the jet shape or profile
function, which describes the angular distribution of energy
with respect to the jet axis [\ref{CDF_D0shape}].
Here one finds a critical point at the (arbitrary) boundary of
the cone used to define the jets, which leads to divergences
in jet shapes and exclusive multijet cross sections at
next-to-leading order and beyond.

In Sect.~\ref{should} we discuss the resummation of these divergences
to all orders. We expect on general grounds that infrared
finiteness will be restored by resummation, and we show that
this is the case for the leading double-logarithmic terms.
After resummation, one obtains a characteristic structure which
is not only finite but smooth (infinitely differentiable)
at the critical point. By analogy with the characteristic
Sudakov peak generated by a similar mechanism near the
exclusive phase space boundary, we call this structure
a {\em Sudakov shoulder}.
We also discuss the resummation of a more general class
of non-smooth behaviours at the critical point, which
give rise to different structures related to the
Sudakov shoulder.

Finally in Sect.~\ref{conc} we summarize our main results and conclusions.

\mysection{Singularities in fixed order}
\label{sing}

The singularities we are interested in are due to the radiation of soft
and/or collinear partons. In perturbative calculations,
the soft and collinear regions of the phase space lead to infrared
divergences in the QCD matrix elements.
In the case of infrared and collinear 
safe observables the divergences cancel upon adding {\em real} and {\em virtual}
contributions but, precisely speaking, the cancellation does not necessarily
take place order by order in perturbation theory. In fact, in particular
kinematic configurations real and virtual contributions can be
highly unbalanced, spoiling the cancellation mechanism.

\noindent {\it Soft-gluon cancellation}

In order to discuss this point most generally, we denote by $C$
an infrared- and collinear-safe observable defined in a hard-scattering
process at the scale $Q$. We can always assume that $C$ is dimensionless
and varies in the kinematic range $0 \leq C \leq 1$. We denote by
${\hat\sigma}^{(n)}(C)$ a related cross section or distribution\footnote{The 
perturbative distribution ${\hat\sigma}(C)$ can be either a measurable quantity
or a short-distance coefficient function that appears in the factorization
theorem.}
evaluated at $n$-th order in perturbation theory and consider the effect
of soft-gluon radiation at the next perturbative order. 
If $1-z$ denotes the fraction of the centre-of-mass energy involved in the
radiative process, virtual $(v)$
and real $(r)$ soft gluons affect the cross section with the following emission
probabilities
\beeq{dwv} 
\frac{dw_v(z)}{dz} &=& - \frac{\as(Q)}{\pi} \;\delta(1-z) \int_0^{1-\epsilon} 
\frac{dz'}{1-z'}
\ln \frac{1}{1-z'} \;, \\
\label{dwr}
\frac{dw_r(z)}{dz} &=& \;\; \frac{\as(Q)}{\pi } \,\frac{1}{1-z} 
\ln \frac{1}{1-z} \; \Theta(1-z-\epsilon) \;.
\eeeq 
The expressions (\ref{dwv},\ref{dwr})
are valid to double-logarithmic accuracy and arise from the combination
of the customary bremsstrahlung spectrum $d\omega/\omega$ with the
angular distribution $d\theta^2/\theta^2$ for collinear radiation.
Here we have introduced an unphysical cutoff $\epsilon$ on the minimum
energy fraction of both gluons because the probabilities in Eqs.~(\ref{dwv}) 
and (\ref{dwr}) are separately divergent if the cutoff is absent.
However, in the case of infrared- and collinear-safe observables,
real and virtual emissions contribute with {\em equal} weight $K$ to cross
sections, although the actual value of the coefficient $K$ depends on
the particular observable under study\footnote{In the case of measurable 
quantities, the coefficient $K$ is typically positive definite. If instead
the distribution ${\hat\sigma}(C)$ is a short-distance coefficient function,
the sign of $K$ depends strongly on the factorization scheme used to
define the parton distribution functions.}.
Thus, adding the real and virtual terms, the physical limit 
$\epsilon \to 0$ can be safely taken, leading to a finite differential
probability:
\beeq{softproep}
\frac{dw(z)}{dz} &=& K \; \lim_{\epsilon \to 0} 
    \, \left[ \frac{dw_v(z)}{dz} + \frac{dw_r(z)}{dz} \right] \\
\label{softpro}   
                 &=&a \left( \frac{1}{1-z} \ln \frac{1}{1-z} \right)_+ \;,
\eeeq
where $a=K \as(Q)/\pi$ and, as usual, the plus-prescription $[ g(z) ]_+$
stands for
\beq{plusd}
\int_0^1 dz f(z) \,[g(z)]_+ \equiv \int_0^1 dz \,[f(z)-f(1)] \,g(z) \;\;.
\eeq

The plus-prescription in Eq.~(\ref{softpro}) is the result of the 
cancellation of the dependence on the unphysical cutoff $\epsilon$ and
defines a well-behaved
distribution, i.e.\ one leading to non-singular functions, when 
acting on any smooth function $f(z)$ that is finite at $z=1$. However,
this is the sole cancellation mechanism of infrared singularities
that is guaranteed by infrared and collinear safety
in fixed-order perturbation theory. The Sterman-Weinberg criteria do
not imply that the coefficients of the perturbative expansion are 
non-singular functions. In fact, normally these 
coefficients are (or are obtained from) singular generalized functions
or distributions, like the soft-gluon probability in Eq.~(\ref{softpro}).
Such distributions lead to finite quantities only when they are integrated
with sufficiently smooth test functions.

In particular, the $(n+1)$-st order cross section ${\hat\sigma}^{(n+1)}(C)$
is obtained by combining the soft-gluon probability with 
the $n$-th order term ${\hat\sigma}^{(n)}(C)$ and the latter 
may well not be a smooth function of $C$ for 
all values of $C$. In this case, the real and virtual
contributions are effectively unbalanced and singularities show up
in the fixed-order perturbative expansion.

In order to discuss these singularities, we have to specify 
how the observable $C$ depends on the soft-gluon energy. We consider
for definiteness the case in which $C$ increases when soft partons are
radiated in the final state.
Thus, the emission of a soft gluon with energy fraction $1-z$ produces a
change $C \to C + \delta C$, where $\delta C = {\cal O}(1-z)$, and, without
loss of generality, we can write the next perturbative order as follows:
\beq{sigmasoft}
{\hat\sigma}^{(n+1)}(C) = \int_0^C dy \; {\hat\sigma}^{(n)}(C-y) 
\;\left(\frac{dw(z)}{dz}\right)_{z=1-y} + \dots \;\;,
\eeq
where the dots stand for less singular terms, if any. Because of the character
of the soft-gluon distribution $dw(z)/dz$, if ${\hat\sigma}^{(n)}$ is not
smooth at some point in the kinematic range $0 \leq C \leq 1$, 
the integral on the right-hand side
of Eq.~(\ref{sigmasoft}) can be divergent.

\vspace{1ex}\noindent {\it Singularities at the exclusive boundary}

The most common and extensively studied soft-gluon singularities are those 
arising at an {\em exclusive boundary} of the phase space. In this kinematic
regime, the radiative tail of real emission is strongly suppressed,  
producing the loss of balance with the virtual contribution.
Since the observable $C$ that we are considering increases
when soft partons are radiated,
the exclusive boundary is the region $C \to 0$. Here 
the non-smooth behaviour of ${\hat\sigma}^{(n+1)}(C)$
is simply due to the associated phase-space constraint $\Theta(C)$.
Inserting Eq.~(\ref{softpro}) into Eq.~(\ref{sigmasoft}) we have:
\beq{sigsoft}
{\hat\sigma}^{(n+1)}(C) = a \left[ - \frac{1}{2} 
{\hat\sigma}^{(n)}(C) \,\ln^2 C 
+ \int_0^C \frac{dy}{y} \ln\frac{1}{y} 
\; \left( {\hat\sigma}^{(n)}(C-y) - {\hat\sigma}^{(n)}(C) \right) \right] 
+ \dots \;\;.
\eeq
Thus, even if ${\hat\sigma}^{(n)}(C)$ is finite\footnote{If
${\hat\sigma}^{(n)}(C)$ is divergent at $C=0$, the double-logarithmic
singularity in Eq.~(\ref{sigsoft0}) is enhanced and can also be non-integrable.}
as $C \to 0$,
the first term in the square bracket produces a divergent
contribution at the kinematical boundary: 
\beq{sigsoft0}
{\hat\sigma}^{(n+1)}(C) 
\Clim - \frac{a}{2} \,{\hat\sigma}^{(n)}(C) \,\ln^2 C + \dots \;\;.
\eeq

These double-logarithmic terms are usually called soft-gluon singularities
of Sudakov type. They appear in the perturbative expansions of many $\ee$
shape variables in the two-jet limit [\ref{CTWT},\ref{shape}]. 
The observable $C$, for instance, can be the \Cpar\ [\ref{Cpar},\ref{CWlett}]
or $C= 1-T$, where $T$ is the thrust [\ref{Farhi}].
Similar soft-gluon effects occur in hadron collisions for 
$Q_\perp$-distributions in the Drell-Yan process ($C \sim Q_\perp/Q$) 
[\ref{QT}] and for the
production of systems of high mass $M$ near threshold ($1-C \sim M/{\sqrt S}$, 
where ${\sqrt S}$ is the centre-of-mass energy). Outstanding examples
of these systems are lepton pairs with large invariant mass produced 
via the Drell-Yan mechanism [\ref{Sterman}],
the hadronic final state in deep-inelastic
lepton-hadron scattering [\ref{CMW}],
heavy quark-antiquark pairs [\ref{sigres}-\ref{CMNT}] and 
pairs of jets at large transverse momentum [\ref{CMNT}].

\vspace{1ex}\noindent {\it Singularities inside the physical region}
 
Another possible source of soft-gluon singularities, which has received
less attention in the literature, arises when ${\hat\sigma}^{(n)}(C)$
has non-smooth behaviour at a certain value $C=C_0$ {\em inside the
physical region} of phase space. 
This can happen if the phase-space
boundary for a certain number of partons lies inside that for a larger
number, or if the observable $C$ is defined in a non-smooth way. We
discuss examples of both types below. 

The diagnosis of singularities inside the physical region is physically less
obvious than that for singularities at the phase-space boundary. At the
exclusive boundary we are dealing with a multi-scale process: besides the hard
scale $Q$, there is also another natural scale, the inelasticity scale 
$Q' \sim C Q$, that plays a
relevant role. The Sudakov singularities in Eq.~(\ref{sigsoft0}) follow from
the large mismatch between these two scales, $Q' \ll Q$ when $C \to 0$. 
For singularities inside the physical region, on the other hand,
the lack of balance between
real and virtual contributions is just produced by the sharpness of the
distribution around the critical point $C_0$,
and the actual identification of a
critical point requires careful analysis of the
kinematics and dynamics.

To show how these singularities arise, we denote the cross
section for $C<C_0$ by ${\hat\sigma}_-(C)$ and that for
$C>C_0$ by ${\hat\sigma}_+(C)$, and we can suppose that at the $n$-th
order both ${\hat\sigma}^{(n)}_-$ and ${\hat\sigma}^{(n)}_+$ are 
infinitely differentiable functions. Then we find from Eq.~(\ref{sigsoft})
that ${\hat\sigma}^{(n+1)}_-(C)$ is infinitely differentiable and regular at 
$C=C_0$ while ${\hat\sigma}^{(n+1)}_+(C)$ can be singular there. The difference
comes from the second term in the square bracket of Eq.~(\ref{sigsoft})
and is given by
\beeq{sigsoftdif}
\!\!\!\!\!\! {\hat\sigma}^{(n+1)}_+(C)-{\hat\sigma}^{(n+1)}_-(C_0)
&\!\!=& - \frac{a}{2} \,\ln^2 (C-C_0) \left\{
\left[{\hat\sigma}^{(n)}_+(C_0)-{\hat\sigma}^{(n)}_-(C_0) \right] \right.
\nonumber \\
&\!\!+& \left. (C-C_0) \left[{\hat\sigma}^{(n)\,\prime}_+(C_0)
-{\hat\sigma}^{(n)\,\prime}_-(C_0) \right] + \ldots \right\} +\ldots \;\;,
\eeeq
where ${\hat\sigma}^{(n)\,\prime}_-(C)$ denotes the first derivative of the
distribution with respect to $C$
and the dots indicate terms that are less singular as $C \to C_0$.
Thus if ${\hat\sigma}^{(n)}_+(C_0)\neq {\hat\sigma}^{(n)}_-(C_0)$,
i.e.\ ${\hat\sigma}^{(n)}(C)$ has a {\em step} at $C=C_0$, then
${\hat\sigma}^{(n+1)}_+(C)$ has a {\em double-logarithmic divergence}
at that point. The same mechanism leading to Eq.~(\ref{sigsoftdif}) will
enhance the double-logarithmic singularity by
further integer powers of $\ln(C-C_0)$ in yet higher
orders\footnote{Note that all these singularities are integrable in a 
neighbourhood of the critical point.}.

In general, a step at some point $C=C_0$ inside the physical region at the
$n$-th order of perturbation theory {\em always} generates more and more 
divergent
contributions at $C=C_0$ in higher and higher orders. To any fixed order
$m>n$, the cross section ${\hat\sigma}^{(m)}_+(C_0)$ is infinite, because the
emission of arbitrarily soft gluons is not cancelled by virtual corrections
at that point.  One may regard this as a failure of the Sterman-Weinberg
criteria for infrared and collinear finiteness in fixed-order perturbation
theory: we have considered an observable which satisfies the criteria,
but nevertheless the perturbative prediction at a point inside the
physical phase-space is divergent.

One might think that the construction of an infrared and collinear safe
quantity, which at the $n$-th perturbative order has a step-like behaviour 
inside the physical region, is quite abstract.
In fact, this can happen if, due to the finite number of partons involved in
the calculation at that order, they can occupy the phase-space region only
up to a value $C=C_0$ that is smaller than the maximum value permitted
by kinematics. To produce a step it is sufficient that the computed 
distribution does not vanish at $C=C_0$.  
As we shall see in Sect.~\ref{exam}, the distribution of the \Cpar\ in
$\ee$ annihilation has these features. 

A similar behaviour can occur if the infrared and collinear safe observable
$C$ is used to classify the topology of hadronic events. In this case sharp
structures can easily be produced. For instance,
$C$ can be a jet resolution parameter that is defined in a non-smooth way
(e.g. it can have a non-trivial and non-smooth dependence on energies and 
angles of partons). Then
the topology of the (partonic) event, e.g. the number of jets, can suddenly 
change at a certain value $C_0$. At that point the rate ${\hat\sigma}^{(n)}(C)$
for producing a fixed number of jets can have a step. Moreover, a corresponding
step, with the opposite discontinuity, will appear in the production rate 
relating to a different number of jets.  At the next perturbative order,
the step in ${\hat\sigma}^{(n)}(C)$ leads to a singularity for $C > C_0$
and, by the same argument as in Eq.~(\ref{sigsoftdif}),
the step in the other jet rate leads to a similar singularity for $C < C_0$.
Since the sum of the two jet rates remains finite, both of them will show
a logarithmic divergence on both sides of the step, that is, {\em above} 
and {\em below} the critical point $C=C_0$. This may appear in 
contradiction with the statement above Eq.~(\ref{sigsoftdif}). However,
since we are considering a variable $C$ that is 
defined in a non-smooth way, it is not monotonic with respect to soft-gluon
energies. Therefore, the simple formula in Eq.~(\ref{sigmasoft})
cannot be applied, although the general relation between steps and
singularities is still valid. A phenomenologically-relevant example
of this kind of singularity is also considered in Sect.~\ref{exam}.

The result in Eq.~(\ref{sigsoftdif}) also allows us to discuss the effect
of soft-gluon radiation on non-smooth quantities when they do not
have a step-like behaviour. For example, the $n$-th-order distribution
${\hat\sigma}^{(n)}(C)$ may be continuous but not its first derivative
(${\hat\sigma}^{(n)\,\prime}_+(C_0)\neq {\hat\sigma}^{(n)\,\prime}_-(C_0)$),
so that ${\hat\sigma}^{(n)}(C)$ has a sharp {\em edge} at $C=C_0$.
An example would be $C=1-T$ where $T$ is the $\ee$ thrust variable,
which has kinematic range $\half \leq T \leq 1$ but is zero in first
order ($n=1$) for $T<\twth$ and vanishes linearly as $T\to T_0=\twth^+$.
Then ${\hat\sigma}^{(n+1)}(C)$ is also continuous but has a {\em cusp} at
$C=C_0$, where ${\hat\sigma}^{(n+1)\,\prime}_+(C)$ diverges
double-logarithmically\footnote{In general, if ${\hat\sigma}^{(n)}$
is continuous up to its $k$-th derivative, the double-logarithmic
singularity affects the  $k$-th derivative of ${\hat\sigma}^{(n+1)}$.}.
In the case of $C=1-T$, from Eq.~(\ref{sigsoftdif}) we obtain that
${\hat\sigma}^{(2)\,\prime}_+(C)$ diverges to negative infinity\footnote{This
cusp is visible in Fig.~1 of Ref.~[\ref{CMHS}].} at $C_0=1-T_0=\thrd$.
In higher orders ${\hat\sigma}(C)$ remains continuous but the degree of
divergence of  its derivative is enhanced by an increasing number of double
logs, so the edge gets more and more like a step order by order in
perturbation theory. 

An edge at a certain order thus does not produce divergences
in the perturbative expansion. Nonetheless, fixed-order
predictions are likely to be unreliable
near the critical point because the shape of the distribution,
independently of its magnitude, is highly unstable with respect to radiative
corrections. For instance, if ${\hat\sigma}^{(n)}(C)$ is decreasing at the
edge and the perturbative coefficient $a$ in Eq.~(\ref{sigsoftdif}) is
positive, the cusp at the next order leads to an increasing distribution
for $C \geq C_0$.

\mysection{Examples}
\label{exam}

Here we illustrate the above general discussion with two specific
examples of distributions of well-known observables which do
develop divergences at points inside the physical phase space,
due to imperfect cancellation between real and virtual soft-gluon
singularities.

\subsection{\boldmath The \Cpar\ in $\ee$ annihilation}
\label{csec}

The \Cpar\ is an infrared- and collinear-safe observable that was introduced 
in Refs.~[\ref{Cpar}] to describe the shape of hadronic events produced
by $\ee$ annihilation. Denoting by ${\bom p}_i^{\alpha}$ the components
of the three-momentum of any final-state  particle $i$, one first considers
the linearized momentum tensor
\beq{ptens}
 \Theta^{\alpha \beta} = \frac{\sum_i {\bom p}^\alpha_i {\bom p}^\beta_i 
/|{\bom p}_i|}
{\sum_j |{\bom p}_j|}\; ,
\eeq
and computes its eigenvalues $\lambda_i$. The \Cpar\ is then defined as
follows
\beq{Cdef}
C = 3 \,(\lambda_1\lambda_2+\lambda_2\lambda_3+\lambda_3\lambda_1) \;\;.
\eeq
By definition the eigenvalues satisfy the following constraints
\beq{const}
0 \leq \lambda_i \leq 1 \;, \;\;\;\;  \sum_i \lambda_i = 1 \;\;,
\eeq
so that the \Cpar\ varies in the kinematic range $0 \leq C \leq 1$.
In particular, $C=0$ for a perfectly two-jet-like
final state (e.g. $ \lambda_2 = \lambda_3 = 0 , \, \lambda_1=1$)
and $C=1$ for an isotropic and acoplanar distribution of 
final-state momenta $(\lambda_1 = \lambda_2 = \lambda_3 = \thrd$).

In fact, the maximal value $C=1$ can only be achieved 
when there are four or more final-state particles. 
Planar events have one vanishing eigenvalue and
occupy the kinematic region $C \leq \thrq$.
Hence, we are in a situation where the phase-space boundary
for three particles is below the kinematic limit of the shape variable. 
For a general three-particle state
\beq{C3def}
C = C_3(x_1,x_2) = 6(1-x_1)(1-x_2)(1-x_3)/(x_1 x_2 x_3)
\eeq
where $x_i=2 p_i\cdot Q/Q^2$ are the centre-of-mass
energy fractions ($x_1+x_2+x_3=2$). The maximum value
$C=\thrq$ for three particles corresponds to the
symmetric configuration $x_1=x_2=x_3=\twth$.

In perturbation theory, the distribution of the \Cpar\ 
for $C\neq 0$
has the general form
\beq{Cexp}
\frac{1}{\sigma_0} \frac{d\sigma}{dC}
 = \asb A(C) + \asb^2 B(C) + {\cal O}(\as^3)\; ,
\eeq
where $\asb=\as(Q)/2\pi$ and we normalize to the Born cross 
section $\sigma_0$, as was done in Ref.~[\ref{KN}]. The two functions
$A(C)$ and $B(C)$ can be identified respectively with the distributions
${\hat \sigma}^{(n)}$ and ${\hat \sigma}^{(n+1)}$ of Sect.~\ref{sing}. In 
particular, we are exactly in the situation where Eq.~(\ref{sigmasoft})
applies because, as can be checked easily, the \Cpar\ increases when additional
soft particles are produced. 

\begin{figure}
  \centerline{
    \setlength{\unitlength}{1cm}
    \begin{picture}(0,7.5)
       \put(0,0){\includegraphics{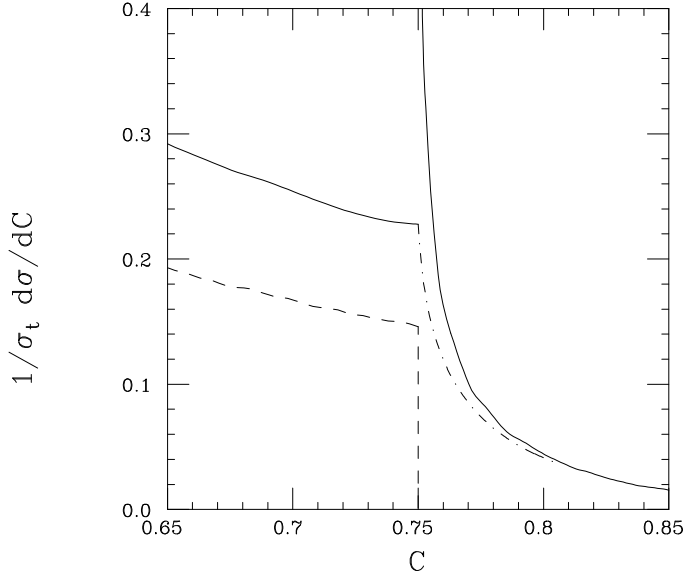}}
    \end{picture}}
  \caption[DATA]{
Predictions of the \Cpar\ distribution for $\as = 0.12$.
Dashed: ${\cal O}(\as)$. Solid: ${\cal O}(\as^2)$.
Dot-dashed: resummed. }\label{fig_Cdist}
\end{figure}

The first-order distribution $A(C)$ 
is given by the $q\bar q g$ final state:
\beq{C3dis}
A(C) = \int_0^1 dx_1 dx_2\, \Theta(x_1 + x_2 -1) \,
M(x_1,x_2)\,\delta[C-C_3(x_1,x_2)]
\eeq
where
\beq{Mx1x2}
M(x_1,x_2) = C_F \, \frac{x_1^2+x_2^2}{(1-x_1)(1-x_2)}\;.
\eeq
Note that the matrix element (\ref{Mx1x2}) has a finite value 
$M_0= M(x_1=x_2= \rat 2 3)=8 C_F$ in the symmetric configuration 
where $C=\thrq$.
Furthermore one finds that the available phase-space remains finite
as $C \to \thrq^-$. Therefore, the 
${\cal O}(\as)$ distribution (\ref{C3dis}) has
a step\footnote{Note that, consistently with the general discussion in 
Sect.~\ref{sing}, the first-order distribution $A(C)$ in Eq.~(\ref{C3dis}) 
diverges at the exclusive boundary $C=0$ of the phase space. This
divergence and the corresponding ones in higher orders can be resummed
[\ref{CWlett}] using the techniques of Ref.~[\ref{CTWT}].}
at the three-parton upper limit $C=\thrq$:
\beeq{ac}
A(C) \!\!\!\!&\Climeq&\!\!\!\! A(\thrq) \;\Theta(\thrq - C) \;\;, \\
\label{C334}
A(\thrq) \!\!\!\!&=&\!\!\!\! \frac{32}{243}\pi\sqrt 3\;M_0 = 
\frac{256}{243}\pi\sqrt 3 \,C_F  \;.
\eeeq
The derivative of $A(C)$ is also finite at 
$C={\thrq}^{-}$:
in fact
\beq{Aprime}
A'(\thrq) = -\rat 8 3 A(\thrq)\; .
\eeq 
The resulting first-order prediction in the vicinity of $C=\thrq$
is shown by the dashed curve in Fig.~\ref{fig_Cdist}.

According to Sect.~\ref{sing}, the second-order contribution $B(C)$ should
be smooth as $C\to \thrq^-$ but should have a double-logarithmic singularity
as $C\to \thrq^+$. This is indeed the case, as shown by the solid curve
in Fig.~\ref{fig_Cdist}. Denoting $B(C)$ above/below $C=\thrq$ by $B_\pm(C)$
we find using the Monte Carlo matrix element evaluation program EVENT
[\ref{KN}]
\beq{B34m}
B_-(\thrq) \;=\; 230\pm 10
\;.
\eeq
Performing the analytic calculation for $C\to \thrq^+$, we find
\beq{Cto34p}
B_+(C)\simeq A(\thrq)\left[ (2C_F + C_A) 
(1-\delta)\ln^2\delta
+(3 C_F+\rat{11}{6} C_A-\rat 1 3 N_f)\ln\delta \right]
+ h(C)
\eeq
where
\beq{epsdef}
\delta= \rat 8 3 \left( C - \rat 3 4 \right)
\eeq
and the remainder is expected to take the form
\beq{d34}
h(C) = h(\thrq) + {\cal O}(\delta \ln\delta)\;.
\eeq
Using Eq.~(\ref{Aprime}), we see that the first term in the square bracket on
the right-hand side of
Eq.~(\ref{Cto34p}) is in complete agreement with Eq.~(\ref{sigsoftdif})
with $a= (2C_F + C_A) \as(Q)/\pi$. The second term in 
the square bracket is due to collinear but non-soft parton splitting:
single-logarithmic contributions of this type were neglected in the simplified
discussion of Sect.~\ref{sing}, which was valid to double-logarithmic
(\DL) accuracy only.

\begin{figure}
  \centerline{
    \setlength{\unitlength}{1cm}
    \begin{picture}(0,7.5)
       \put(0,0){\includegraphics{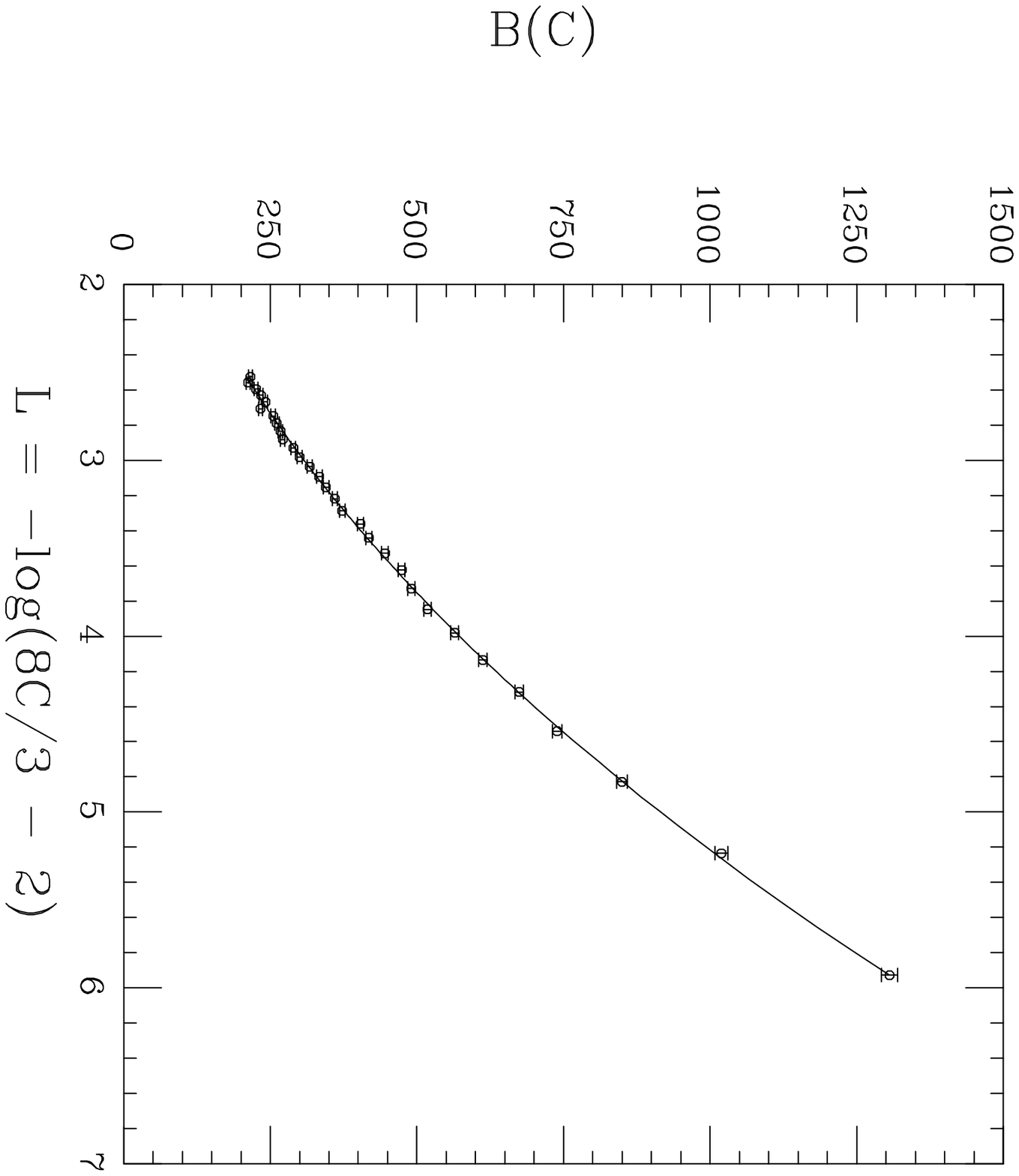}}
    \end{picture}}
  \caption[DATA]{
Second-order prediction of the \Cpar\ distribution for 
$C\to \thrq^+$.\\
Points: \EV\ \MC. Curve: Eqs.~(\ref{Cto34p}-\ref{d0}).
 }\label{fig_Ccoef}
\end{figure}

Comparing our analytic calculation with numerical data from EVENT [\ref{KN}],
we find good agreement, as shown in Fig.~\ref{fig_Ccoef}. The value of
the constant $h(\thrq)$ in Eq.~(\ref{d34}) is found to be
\beq{d0}
h(\thrq) \;=\; 146\pm 3 \; .
\eeq

\subsection{Jet shapes in hadron collisions}
\label{jets}

As our second example we study the angular distribution of
transverse energy ($E_T$) in jets produced at high $E_T$ in hadron-hadron
collisions [\ref{CDF_D0shape}]. Here it is customary to define a jet by
a cone algorithm [\ref{snowmass}], that is, by
maximizing $E_T$ with respect to the direction of a cone of opening
angle $R$ in pseudorapidity, azimuth ($\eta,\phi$) space.
The cone axis then defines the jet axis ($\eta_J,\phi_J$).
Alternatively a $\kper$-clustering algorithm can be
used [\ref{kpalg}]. In this case again a parameter $R$ can be
introduced, effectively representing the jet cone size [\ref{EllSop}].

Several different versions of the cone algorithm are actually used both in
theoretical calculations and in experimental analyses.
They differ with regard to minor details and sometimes major points.
The latter are related to infrared and collinear safety.
In some versions [\ref{UA2}],
the direction of the unassigned cluster (parton) with the highest $E_T$
is used to start the clustering procedure.
Since the highest-$E_T$ parton is not stable with respect to collinear 
splitting, these algorithms are {\em collinear} unsafe.
Their unsafeness certainly affects any perturbative 
calculation that involves more than 5 partons in the final state. 
Collinear safety is fulfilled by iterative-cone algorithms that use all 
clusters, possibly excluding those with $E_T$ below a fixed threshold value 
[\ref{CDF_D0}], as seed directions of the cone. Nonetheless, these
algorithms are still {\em infrared} unsafe in the absence of an
$E_T$-threshold [\ref{KilGie}], and therefore strongly affected
by the threshold when one is applied [\ref{Seymour}].
In iterative-cone algorithms, infrared and collinear safety
can be achieved [\ref{EKS}] by using the $E_T$-weighted 
midpoints of all pairs of jets (partons) as additional seed directions. 
In the following, by `cone algorithm' we refer to these improved versions. 

We consider here the differential jet shape or profile function $\psi(r)$,
defined such that $\psi(r)\,dr$ is the fraction of transverse energy
lying within a ring of radius $r$, width $dr$, centred on the jet axis. We have
\beq{psidef}
\psi(r) = \frac{\sum_i \,E_{Ti} \;\delta(r-R_{iJ})}{\sum_k \,E_{Tk}
\Theta(R-R_{kJ})} \;\;,
\eeq
where
\beq{RiJdef}
R_{iJ} \equiv \sqrt{(\eta_{i}-\eta_{J})^2+(\phi_{i}-\phi_{J})^2}
\eeq
and the sums over $i,k$ can be over either all particles or only those
particles assigned to the jet.  
The function $\psi(r)$ has been studied in detail in Ref.~[\ref{Seymour}],
and many of the points we make are also discussed there.

For $r > 0$, $\psi(r)$ starts at order $\as$
and at this order its $r$-dependence is due to 3-parton final-states.
Let us simplify the discussion by considering a cone size $R < \pi/3$
and three partons of momenta $p_1,p_2,p_3$ with transverse-energy ordering,
e.g.\ $E_{T1} > E_{T2} > E_{T3}$. In this case there are two possible jet
configurations:

a) jets = $\{1\} , \{2\} ,  \{3\} \;\;\; \;\;\mbox{if}
\;\;\; R^{alg}_{23} > R$

b) jets = $\{1\} , \{23\} \;\; \;\;\; \;\;\;\;\;\;\mbox{if}
\;\;\, R^{alg}_{23} < R$

\noindent where $R^{alg}_{23}$ depends on the jet-algorithm:
\beq{R23} 
R^{cone}_{23} = \frac{E_{T2}}{E_{T2} + E_{T3}} R_{23}\;,
\;\;\; \;\;R^{\kper}_{23}= R_{23}\;.
\eeq
The jet shape $\psi(r)$ can be written
in terms of two contributions $\psi_{in}(r)$ and $\psi_{out}(r)$, 
arising from particles assigned and not assigned to the jet, respectively.
To order $\as$, the two contributions are proportional to the following
expressions
\beeq{psim}
\psi_{in}^{(1)}(r) &\propto& \Theta(R - R^{alg}_{23}) 
\; \delta( J - \{23\} ) \nonumber \\
&\cdot& 
\left[ E_{T2} \;\delta\left(r - \frac{E_{T3}}{E_{T2} + E_{T3}} R_{23}\right)
+ E_{T3} \;\delta\left(r - \frac{E_{T2}}{E_{T2} + E_{T3}} R_{23}\right)
\right] \;\;,
\eeeq
\beq{psip}
\psi_{out}^{(1)}(r) \propto \Theta(R^{alg}_{23} - R) \left[ \delta( J - \{2\} ) 
\;E_{T3}
+ \delta( J - \{3\} ) \;E_{T2} \right] \; \delta(r - R_{23}) \;\;,
\eeq
where we use the notation
\beq{deltaJ}
\delta( J - \{2\} ) \equiv 
\delta( E_{TJ} - E_{T2}) \,\delta( \eta_{J} - \eta_{2}) 
\,\delta( \phi_{J} - \phi_{2})
\eeq
and likewise for $\delta( J - \{3\} )$ and $\delta( J - \{23\} )$.
Equations~(\ref{psim}) and (\ref{psip}) show that
$\psi_{in}^{(1)}(r)$ and $\psi_{out}^{(1)}(r)$ are respectively proportional to
$\Theta(R-r)$ and $\Theta(r-R)$, and when
$r \rightarrow R$  we have
\beeq{psi1cone}
\psi_{in}^{(1)}(r) &\to& \psi_{in}^{(1)}(R) > 0 \,,
\;\;\; \psi_{out}^{(1)}(r) \to 0 \,, 
\;\;\; \;\;\; \;\;\; \;\;\; \;\;\;\;(\mbox{cone alg.}) \\
\label{psi1kt}
\psi_{in}^{(1)}(r) &\to& 0 \,, \;\;\; \;\;\; \;\;\; \;\;\; \;\;\;\; 
\;\;\;\psi_{out}^{(1)}(r) 
\to \psi_{out}^{(1)}(R) > 0 \;
\;\;\;(\kper \;\mbox{alg.})
\eeeq
In conclusion,  $\psi^{(1)}(r)$ in the cone algorithm has a step
at $r=R$, while in the $\kper$ algorithm it has an edge (step) if
only particles assigned to the jet (all particles) are considered.

Let us now investigate possible singularities at the next order. Since
$\psi^{(2)}(r) \propto \sum_i E_{Ti}$ \newline $\delta(r - R_{iJ})$, 
we should consider how the angular distance
between the jet direction and that of parton $i$
is affected by collinear splitting and soft gluon radiation.
In the case of collinear splitting the angular distance
does not change. Thus we have to examine soft-gluon radiation
with energy fraction $1-z$ and single-logarithmic emission probability 
$\frac{dw(z)}{dz} = a \left( \frac{1}{1-z} \right)_+$.
 
In the cone algorithm, $R_{iJ}$ is always smaller than $R$ if $i$
belongs to the jet and always larger than $R$ if it does not.
If $i$ belongs to the jet and a soft gluon is emitted inside the
cone, cancellation of real and virtual divergences will occur as
usual, leading to a smooth contribution to $\psi^{(2)}_{in}$.
However, if the gluon is emitted outside the cone there is a
mismatch of real and virtual contributions, which will generate a
divergence at the boundary of the cone.  The corresponding change
in $R_{iJ}$ is $-c(1-z)$, where $c>0$ depends on the kinematics.
Similarly, if parton $i$ is outside the jet and a gluon is emitted
into the cone, the shift in $R_{iJ}$ is $+c(1-z)$. Thus, according
to our general discussion in Sect.~\ref{sing}, we find
\beeq{psiin}
\psi_{in}^{(2)}(r) &\simeq& \Theta(R-r) \; \left\{ -a \; \psi_{in}^{(1)}(R) \;
\ln \frac{1}{R-r} + \const\right\} \nonumber \\
\psi_{out}^{(2)}(r) &\simeq& \Theta(r-R) \;\const
\eeeq
The coefficient $a$ is likely to be positive in most jet configurations
so that $\psi^{(2)}(r)$ diverges to $-\infty$ when $r \to R$ from below
and stays finite when $r \to R$ from above.
 
In the $\kper$ algorithm at higher orders, 
because of the iterative recombination procedure, $R_{iJ}$
can be either larger or smaller than $R$, independently of
whether $i$ is or is not a particle in the jet. Thus both 
$\psi_{in}^{(2)}(r)$ and $\psi_{out}^{(2)}(r)$ are non-vanishing in a
neighbourhood of $r=R$. Moreover, $R_{iJ}$ does not vary
monotonically with $z$ when a soft gluon is added to the
jet. Thus we find that $\psi_{in}^{(2)}(r)$ has a cusp at $r=R$
and $\psi_{out}^{(2)}(r)$ has the following {\em double-sided} singular 
behaviour\footnote{Assuming $a >0$, this agrees with the results of
Ref.~[\ref{Seymour}].}
\beq{psiout}
\psi_{out}^{(2)}(r) \simeq a \left\{ \Theta(R-r) \;\psi_{out}^{(1)}(R) \;
\ln \frac{1}{R-r} - \Theta(r-R) \;\psi_{out}^{(1)}(R) \;\ln \frac{1}{r-R}
+ \const \right\} \;.
\eeq
The double-sided singularity follows from our general discussion in 
Sect.~\ref{sing} on non-monotonic observables with stepwise 
behaviour at lower order.

Note that the singularities in Eqs.~(\ref{psiin}) and (\ref{psiout})
are both single-logarithmic. This is because they are due to soft but
non-collinear radiation.

Related problems connected with the calculation of jet cross sections
in higher orders have been discussed in Ref.~[\ref{KilGie}].
The definitions of the fixed-cone and iterative-cone algorithms
given there are actually unsafe beyond a certain order.
More precisely, since the highest unassigned  $E_T$-cluster (parton) is
used to start the clustering procedure, these algorithms
are {\em collinear} unsafe. For a cone size $R < \pi/3$,
their unsafeness affects perturbative  calculations that
involves more than 5 final-state partons. Thus one may still
perform calculations up to order $\as^4$ using these definitions,   
provided that divergences at critical points are carefully handled.

Consider the 3-jet cross section in the iterative cone algorithm as
defined in Ref.~[\ref{KilGie}] at a fixed value of $R_{23}$, the
angular distance between the two less energetic jets.
At \LO\ this cross section has a step at the critical point $R_{23}=R$,
the cone radius. Actually it vanishes for $R_{23} < R$ and is finite
and not vanishing for  $R_{23} > R$. 
At \NLO\ the emission of a soft gluon at a large angle produces 
a single-logarithmic divergence (the soft gluon not being collinear)
at $R_{23}=R$, of the type discussed above.

This divergence is integrable and it does not produce any
singularity in the cross section integrated over $R_{23}$.
On the other hand, if the phase-space slicing method [\ref{BaOhOw}]
is used to perform the calculation, the integrable divergence can give rise 
to a $\ln\smin$ dependence, where $\smin$ is the `slicing parameter' used for
the numerical evaluation of the cross section.
Such a dependence can be numerically important, even though it should
disappear for sufficiently small values of $\smin$. 

To illustrate this point we note that 
the cutoff $\epsilon$ introduced in Sect.~\ref{sing}
to define the plus-distribution is somewhat analogous to a
slicing parameter. In the case of a stepwise behaviour of
${\hat \sigma}^{(n)}$ at $C=C_0$, the integral up to $C_1 > C_0$,
\beq{limeps}
\lim_{\epsilon \to 0} \int_{C_0}^{C_1} dC \;
{\hat \sigma}^{(n+1)}(C;\epsilon)\;,
\eeq
is in fact finite, but if we keep $\epsilon$ fixed we get for
$C_1 - C_0<\epsilon$
\beq{intC}
\int_{C_0}^{C_1} dC \; {\hat \sigma}^{(n+1)}(C;\epsilon) =
a \left[ - \frac{1}{2} \ln^2 \epsilon \int_{C_0}^{C_1} dC
{\hat \sigma}^{(n)}_+(C) + \const
+ {\cal O}(\epsilon \ln \epsilon) \right]   \;,
\eeq
which shows an unphysical dependence on $\ln \epsilon$.
The unphysical logarithmic dependence does
not cancel as long as $\epsilon$ is not sufficiently small.

\mysection{Resummation and the Sudakov shoulder}
\label{should}

Although we have shown that divergent fixed-order predictions
will generally arise at non-smooth points inside the
physical region, one might feel that this is a purely
academic difficulty, since the divergence at such a point
$C=C_0$ is integrable, as long as
${\hat\sigma}^{(n)}(C)$ itself is integrable in a neighbourhood of
that point. The cancellation of real and virtual divergences will
therefore take place in the presence of smearing in $C$, as would be
expected from non-perturbative (hadronization) effects. However, the
resulting prediction will be highly unstable with respect to the order
at which we stop the perturbative calculation and the amount of
smearing introduced.

In particular, since the hadronization smearing should cancel divergent terms
proportional to some power of $\as(Q)$, this would imply that non-perturbative
effects scale logarithmically with $Q$, thus spoiling not only the {\em
finiteness} but also the {\em safety} of the Sterman-Weinberg criteria. We
should then be forced to
conclude that some infrared- and collinear-safe observables are
affected by non-perturbative contributions that are not power suppressed.

Alternatively, one might think that the divergence in the fixed-order
prediction can be cancelled by the resummation of still higher-order
perturbative corrections. One might argue that the resummation of real
and virtual terms produces a Sudakov suppression of the divergence,
similarly to the Sudakov suppression of many observables at the exclusive
boundary of the phase space [\ref{CTWT}-\ref{QT}]. This scenario would
lead to a resummed prediction that, at the critical point $C=C_0$, is
finite but still has a stepwise behaviour. On physical grounds we do not
anticipate this behaviour, and we should expect non-perturbative effects
to fill up the gap between the two sides of the step. Since the gap is
produced in perturbation theory, in this case also the necessary
non-perturbative contributions could not be power suppressed.

We believe that instead the problem of divergences inside
the physical region has a satisfactory solution
entirely within the context of perturbation theory. Namely, the
resummation of the soft-gluon contributions to all orders, rather than
producing a Sudakov suppression of the divergences, leads to a
structure that is continuous and smooth, indeed infinitely differentiable,
at $C=C_0$. We call this structure a {\em Sudakov shoulder}. The
existence of a Sudakov shoulder implies the restoration of validity
of the Sterman-Weinberg criteria at infinite order in perturbation theory.

To demonstrate this point we first recall how soft-gluon resummation produces
the Sudakov form factor at the exclusive boundary of the phase space. We thus 
consider the case in which ${\hat \sigma}(C)$ is a measurable quantity
(with a positive soft-gluon coefficient $a$) and proceed to
the iterative application of Eq.~(\ref{sigsoft}), 
retaining for simplicity only the double-logarithmic
(\DL) contribution at each order. Taking into account the symmetrization
with respect to exchange of soft gluons, we obtain an exponential
series that, after summation in the neighborhood of the point $C=0$, gives
\beq{siginf0}
{\hat\sigma}^{(\infty)}(C) 
{\raisebox{-1ex}{\rlap {\tiny $\;\;C\to 0$}} 
\raisebox{0ex}
{$\;\;\;\,=\;\;\;\,$}}
 \exp\left\{ - \frac{a}{2} \,\ln^2 C \right\}
\; {\hat\sigma}^{(n_0)}(C) \;\;,
\eeq
where $n_0$ represents the lowest perturbative order for our observable. The
exponential term on the right-hand side of Eq.~(\ref{siginf0}) is the customary
Sudakov form factor, which suppresses the observable at the exclusive boundary.

The analogous iterative procedure in the region $C \sim C_0$ also leads to
an exponential series which can be summed to give
\beq{siginfdif}
{\hat\sigma}^{(\infty)}_+(C)-{\hat\sigma}^{(\infty)}_-(C_0)
=\exp\left\{ - \frac{a}{2} \,\ln^2 (C-C_0) \right\}
\left[{\hat\sigma}^{(n_0)}_+(C_0)-{\hat\sigma}^{(n_0)}_-(C_0) \right] \;,
\eeq
where $n_0$ represents the order at which the step at $C=C_0$ first
appears.  Now we may assume that ${\hat\sigma}^{(\infty)}_-(C_0)$
has no soft-gluon singularities, since by hypothesis $C$ is always
increased by soft gluon emission. It then follows from
Eq.~({\ref{siginfdif}) that after resummation the cross section 
is finite, continuous and infinitely differentiable at $C=C_0$.
Rather than suppressing the divergence at $C > C_0$, Sudakov
resummation leads to the suppression of the step.
The general form of the resummed cross section will be a smooth
extrapolation from ${\hat\sigma}^{(\infty)}_-(C_0)$ into
the region $C>C_0$, joining smoothly with the different
asymptotic behaviour of ${\hat\sigma}^{(\infty)}_+(C)$ well
above $C=C_0$. This is the characteristic structure we call
a Sudakov shoulder.

The dot-dashed curve in Fig.~\ref{fig_Cdist} illustrates the
Sudakov shoulder in the case of the \Cpar\ distribution.
Here we simplify the treatment of the region $C >\thrq$
not only by using the \DL\ approximation for the form factor
but also by approximating
the lowest-order contribution $A(C)$ in Eq.~(\ref{C3dis})
by a step function, $A(C) \simeq A(\thrq) \Theta(\thrq - C)$.
We then obtain
\beq{DLsigma34}
\frac{1}{\sigma} \frac{d\sigma}{dC} \simeq \asb A(\thrq) 
\left\{  \Theta(\thrq - C) + \Theta(C - \thrq) \left(1 -
\exp \left[ - 2 A^{(1)} \asb \ln^2 (C -  \thrq) \right] \right) \right\} \; 
\;\;\;\;\;\;\;\mbox{(\DL)} \;,
\eeq
where
\beq{a1coef}
A^{(1)} = C_F + \half C_A \;.
\eeq
Expanding the right-hand side of Eq.~(\ref{DLsigma34}) perturbatively,
one finds a series of logarithmically divergent (although integrable)
terms for $C \to {\thrq}^+$.
The summation of these terms to all orders leads to a finite result and
the lowest-order step-like behaviour is smoothed. The resummed expression
(\ref{DLsigma34}) shows a shoulder that extends beyond the value $C=\thrq$.
The shoulder becomes less steep with increasing $A^{(1)}$ or $\as$.
Since $A^{(1)}=C_F + C_A/2 \sim 2C_F$,
in an abelian theory (i.e., setting $C_A=0$) the \DL\ shoulder would be
twice as steep as in QCD. The additional broadening in the QCD case is 
due to the fragmentation of the gluon jet. Note that the shoulder is
nevertheless still quite sharp on the scale shown in Fig.~1.

Soft-gluon resummation will smooth out perturbative predictions of
infrared- and collinear-safe observables at other types of critical
points inside phase space. After the case of a step, the next simplest
example is that of a critical point $C=C_0$ with an edge at a given
perturbative order $n_0$.  The resummed behaviour at this point is
easily obtained by noting that the first derivative
${\hat \sigma}^{\prime}(C)$ of the corresponding distribution is 
a safe quantity that has precisely a step at order $n_0$. Thus we can
apply to ${\hat \sigma}^{\prime}(C)$ a DL resummation analogous to that carried
out in Eq.~(\ref{siginfdif}),
to obtain the following distribution
\beq{siginfdifedge}
{\hat\sigma}^{(\infty)}_+(C) = {\hat\sigma}^{(\infty)}_-(C_0)
+ (C-C_0){\hat\sigma}^{(\infty) \,\prime}_-(C_0) +
\left[{\hat\sigma}^{(n_0) \,\prime}_+(C_0) - {\hat\sigma}^{(n_0)
\,\prime}_-(C_0) \right] \int_0^{C-C_0} dx e^{- \frac{a}{2} \ln^2x} \;.
\eeq 
The first two terms on the right-hand side extrapolate   
${\hat\sigma}^{(\infty)}_-(C)$ linearly into the region above the critical
point $C=C_0$ and the last term bends this linear extrapolation
smoothly, in proportion to the sharpness of the lower-order edge. 

The resummation of double-sided singularities of the type encountered
in jet profiles is technically more complicated. 
It leads to a smooth ``jump'' structure 
obtained by smearing (convoluting) two Sudakov shoulders: a shoulder 
above and an inverted shoulder below the critical point, or vice versa. 
As a result, the two sides of the lower-order step match
with some intermediate 
value at the critical point in the all-order distribution and the step 
is smoothed. An example of this resummation carried out numerically for the 
jet shape $\psi(r)$ is presented in Ref.~[\ref{Seymour}] (cf. Fig.~11).
In general, the derivation of resummed expressions in analytic form is 
quite difficult and not always feasible. This is because the type of smearing
(convolution) to be applied to the two shoulders is strongly dependent
on the detailed kinematics. In the presence of double-sided singularities,
the kinematics is usually very involved since it is responsible for the
non-monotonic behaviour of the observable with respect to variations of 
particle angles and energies.

\mysection{Discussion and conclusions}
\label{conc}

In this paper we have discussed in rather general terms the divergences
which can arise inside the physical region in infrared- and collinear-safe
quantities computed to any fixed order in perturbation theory.
We have also given specific examples of widely-used quantities in
$\ee$ and hadron-hadron physics for which these problems actually
arise. The divergences correspond to integrable singularities and
therefore they could in principle be removed by non-perturbative
smearing effects. However, this would require non-perturbative
contributions that are not power-suppressed at high energies.

We have argued that instead the remedy for these problems lies
entirely within perturbation theory: the resummation of soft-gluon
contributions to all orders should be sufficient to
render any safe quantity finite and smooth throughout the
physical phase space. 
Despite the violation of finiteness of fixed-order calculations,
perturbative resummation suggests that non-perturbative effects
are still power-suppressed in infrared- and collinear-safe observables.

A resummation of double-logarithmic
terms was performed to illustrates how this works in the
case of the $\ee$ \Cpar\ distribution (Fig.~1). More work
remains to be done in order to resum single-logarithmic
terms and the type of double-sided singularities encountered
in hadronic-jet profiles.

The issues we have discussed are of some importance for QCD
phenomenology.  Before using any ``safe'' observable to test the
theory or to measure $\as$, one needs to identify the critical
points of that observable and the expected behaviour in whatever
order of perturbation theory is to be used.  These points will
need to be avoided in comparisons between fixed-order predictions
and experiment. On the other hand, if resummed predictions can
be obtained to single-logarithmic precision, then the behaviour
at critical points can be used to provide interesting new
tests of QCD, as was the case for resummation near the 
exclusive phase-space boundary [\ref{CTWT},\ref{shape}].

\section*{Acknowledgments}
We are grateful for helpful discussions on this topic with M.H.\ Seymour,
and for correspondence with W.T.\ Giele.

\section*{References}
\begin{enumerate}
\item \label{book}
  See, for instance: 
  A.\ Bassetto, M.\ Ciafaloni and G.\ Marchesini, \prep{100}{201}{83}; 
  Yu.L.\ Dokshitzer, V.A.\ Khoze, A.H.\ Mueller and S.I.\ Troyan, 
  {\em Basics of Perturbative QCD}\/ (Editions Frontieres, Paris, 1991); 
  R.K.\ Ellis, W.J.\ Stirling and B.R.\ Webber, {\em QCD and Collider
  Physics} (Cambridge Univ. Pr., Cambridge, 1996).
\item \label{SW}
  G.\ Sterman and S.\ Weinberg, \prl{39}{1436}{77}.
\item \label{CSS}
  J.C.\ Collins, D.E.\ Soper and G.\ Sterman, in {\em Perturbative Quantum
  Chromodynamics}, ed. A.H.\ Mueller (World Scientific, Singapore, 1989), 
  p.~1 and references therein.
\item \label{CTWT}
  S.\ Catani, G.\ Turnock, B.R.\ Webber and L.\ Trentadue, \np{407}{3}{93}.
\item \label{shape}
  S.\ Catani, G.\ Turnock, B.R.\ Webber and L.\ Trentadue, \pl{263}{491}{91};
  S.\ Catani, G.\ Turnock, B.R.\ Webber, \pl{272}{368}{91}, 
  \pl{295}{269}{92}.
\item \label{QT} 
  J.\ Kodaira and L.\ Trentadue, \pl{112}{66}{82}, \pl{123}{335}{82}; 
  C.T.H.\ Davies,  W.J.\ Stirling and B.R.\ Webber, \np{256}{413}{85};
  J.C.\ Collins, D.E.\ Soper and G. Sterman, \np{250}{199}{85}.
\item \label{Sterman}
  G.\ Sterman, \np{281}{310}{87};
  S.\ Catani and L.\ Trentadue, \np{327}{323}{89}, \np{353}{183}{91}. 
\item \label{CMW}  
  S.\ Catani, G. Marchesini and B.R.\ Webber, \np{349}{635}{91}.
\item \label{sigres}
  E.\ Laenen, J.\ Smith and W.L.\ van Neerven, \np{369}{543}{92}; 
\item \label{BC}
  E.L.\ Berger and H.\ Contopanagos, \pr{54}{3085}{96}.
\item \label{CMNT}
  S.\ Catani, M.L.\ Mangano, P.\ Nason and L.\ Trentadue, 
  \np{478}{273}{96}.
\item  \label{Cpar}
  G.\ Parisi, \pl{74}{65}{78}; \\
  J.F.\ Donoghue, F.E.\ Low and S.Y.\ Pi, \pr{20}{2759}{79}; \\
  R.K.\ Ellis, D.A.\ Ross and A.E.\ Terrano, \np{178}{421}{81}.
\item \label{CWlett}
  S.\ Catani and B.R. Webber, \cav{97/16}, in preparation.
\item  \label{CDF_D0shape}
  CDF Collaboration, F.\ Abe et al., \pr{70}{713}{93};
  D0 Collaboration, S.~Abachi et al., \pl{357}{500}{95}.
\item  \label{Farhi}
  E.\ Farhi, \prl{39}{1587}{77}.
\item  \label{CMHS}       
  S.\ Catani and M.H.\ Seymour, \pl{378}{287}{96}.
\item  \label{KN}
  Z.\ Kunszt, P.\ Nason, G.\ Marchesini and B.R.\ Webber, in
  `Z Physics at LEP 1', CERN 89-08, vol.~1, p.~373.
\item \label{snowmass}
  S.D.\ Ellis et al., in Proceedings of {\it Research Directions for the
  Decade, Snowmass 1990}, ed. E.L.\ Berger (World Scientific, Singapore,
  1992), p.~134. 
\item  \label{kpalg}
  S.\ Catani, Yu.L.\ Dokshitzer, M.H.\ Seymour and B.R.\ Webber,
  \np{406}{187}{93}.
\item  \label{EllSop}
  S.D.\ Ellis and D.E.\ Soper, \pr{48}{3160}{93}.
\item  \label{UA2} 
  UA2 Collaboration, J.\ Alitti et al., \pl{257}{232}{91}. 
\item  \label{CDF_D0}
  CDF Collaboration, F.\ Abe et al., \pr{45}{1448}{92};\\
  D0 Collaboration, S.\ Abachi et al., \pr{53}{6000}{96}.            
\item  \label{KilGie}
  W.B.\ Kilgore and W.T.\ Giele, \pr{55}{7183}{97};\\
  W.B.\ Kilgore, preprint Fermilab-Conf-97/141-T (hep-ph/9705384). 
\item  \label{Seymour}
  M.H.\ Seymour, preprint RAL-97-026 (hep-ph/9707338).
\item  \label{EKS}
  S.D.\ Ellis, Z.\ Kunszt and D.E.\ Soper, \pr{62}{726}{89}.
\item  \label{BaOhOw}
  H.\ Baer, J.\ Ohnemus and J.F.\ Owens, \pr{40}{2844}{89}.
\end{enumerate}
\end{document}